# Quantitative approximation of geothermal potential of Bakreswar geothermal area in eastern India

Chiranjit Maji[*], Saroj Khutia, Hirok Chaudhuri

Department of Physics, National Institute of Technology Durgapur, Mahatma Gandhi Avenue, Durgapur- 713209, India.

[*] Corresponding author's email:   chiranjit.nitdphysics@yahoo.in;
cm.15ph1102@phd.nitdgp.ac.in

**Abstract**

Proper utilization of geothermal energy for power generation is still ignored in India even after having it's enough potential as much as the equivalent to the other nonconventional energy resources of the country. A major thrust is required in this field of technology. The source of geothermal energy is the decay of the radio-nuclei such as Uranium, Thorium, and Potassium inside the Earth's crust apart from the primordial heat source. The noble gas $^4$He is also produced during the radioactive disintegration process. Therefore, measuring the amount of $^4$He gas generated in the terrestrial radioactive process along with some other geochemical parameters in an Indian geothermal reservoir, the potential of the reservoir can be evaluated without performing conventional and detailed geochemical & geophysical techniques. Mathematical calculations relating to the radioactive disintegration to estimate the geothermal potential of the Bakreswar geothermal area at eastern India utilizing the concept of the $^4$He exploration technique has been described here. The study showed that the heat energy generated by the radioactive decay of $^{238}$Th, $^{238}$U and $^{235}$U inside Bakreswar geothermal reservoir was evaluated as 38 MW only considering the He emanated from the Agni Kunda hot spring. In addition, the depth of the geothermal reservoir was also evaluated to be about 1,100 m using some geophysical characteristics of the reservoir at the study area. Furthermore, the suitable locations of deep drilling for the installation of the probable geothermal power plant are also identified by investigating the resistivity survey profile of the study area.

1.  **Introduction**

   Geothermal energy is the heat generated due to the decay of the radio–nuclei inside the Earth's crust in addition to the primordial heat source which was trapped inside the Earth's interior during the accretion process of our planet. The origin of this heat is linked with the internal structure of our planet and the physical processes occurring there. Geothermal energy, which is found in the Earth's crust, is unevenly distributed throughout the globe at near the surface to the deep Earth up to 16 km or even more (Barbier, 2002; Billings, 1990). This natural heat energy is being exploited throughout the globe for the commercial production of electric power using modern technology (Breeze, 2019; Mburu, 2009). The geothermal gradient which is the rate of the increment of temperature profile of the Earth's underneath bedrock, has a global average value of 30 °C/km in continental crust (Barbier, 2002) and 100 °C/km in oceanic crust (Cann, 1979) however, in geothermal areas, the values of the said gradient are well above (>40 °C/km) (Armstead, 1983) the global average value (30 °C/km). It happens when molten magma moves upwards in a special circumstance through joints and fractures, and it is trapped within the Earth's crust at a depth of 5–10 km beneath the Earth's surface and forms a magmatic intrusion in the form of igneous (created under intense heat) rock (Billings, 1990; Csányi et al., 2010; Dickson and Fanelli, 2003). This intrusion may still in a fluid state or the process of solidification through cooling and releasing heat constantly. The origin of geothermal energy is categorized into two sources. One of these was from a relic of the planet's accretion process when the Earth's internal temperature was much higher than it is now and huge energy was trapped within the Earth at about 4.5 billion years ago (Csányi et al., 2010). This is known as the primordial heat source. Another one is known as the radiogenic heat source which is produced by the decay of long-lived radioisotopes such as $^{238}$U, $^{235}$U, $^{232}$Th and $^{40}$K which are having a half-life ($T_{1/2}$) comparable to the age of the Earth and found within the Earth's crust with sufficient abundance in geothermal areas (Beus, 1976; Pereira, 1980). For example, in the Earth's crust, $^{238}$U [or $^{235}$U] and $^{232}$Th are present in the range of 0.1 to 19.7 ppm and 0.1 to 56.0 ppm respectively (Pereira, 1980) whereas the estimated crustal abundance of $^{40}$K varies within 2200 to 51100 ppm (Beus, 1976). Therefore, these radioactive isotopes are treated as a considerable source of heat in the deep Earth. The physico-chemical processes and the heat energy produced from the naturally occurring radioactive isotopes are shown below (Chaudhuri et al., 2019; Cupps, 2015). Moreover, crustal He ($^4$He) atoms and neutrinos are also produced during the decay process as described below.

$$^{232}_{90}\text{Th} \rightarrow {}^{208}_{82}\text{Pb} + 6\,{}^{4}_{2}\text{He} + 4\,{}^{0}_{-1}\text{e} + 42.60 \text{ Mev/atom} \quad (1)$$

$$^{238}_{92}\text{U} \rightarrow {}^{206}_{82}\text{Pb} + 8\,{}^{4}_{2}\text{He} + 6\,{}^{0}_{-1}\text{e} + 51.70 \text{ Mev/atom} \quad (2)$$

$$^{235}_{92}\text{U} \rightarrow {}^{207}_{82}\text{Pb} + 7\,{}^{4}_{2}\text{He} + 4\,{}^{0}_{-1}\text{e} + 46.40 \text{ Mev/atom} \quad (3)$$

Moreover, the literature survey invokes that the production rate of He from $^{238}$U [and $^{235}$U] and $^{232}$Th isotopes within the deep Earth to be as $1.03 \times 10^8\ atoms/m^3/s$ and $2.43 \times 10^{10}\ atoms/m^3/s$, respectively (Pereira, 1980). Such physico–chemical processes suggest the co-existence of unique heat–helium coherence under the deep Earth at any geothermal system (Kennedy et al., 2000) and this radiogenic heat is the main source of the Earth's internal heat, which in turn, powers all geodynamic processes (Joshua et al., 2012). In general, geothermal heat from the aquifer (reservoirs) is transferred to the Earth's surface by the process of conduction and convection and geothermal fluid i.e. meteoric water acts as the carrier (Barbier, 2002) and the radiogenic He, being highly diffusive gas, produced from host mineral mixes with the fluid that circulates into the deep Earth by diffusion (Nicholson, 1993). The reservoir is a volume consisting of hot permeable rocks which is generally sandwiched by a capping of impermeable rocks and connected to a surficial recharge area (Hochstein, 1990). Usually, geothermal fluids percolated into the deep Earth and recharged the aquifer cyclically (Chaudhuri et al., 2018; Hochstein, 1990) and heat is transferred from the reservoir to the circulating fluids which escape from the deep reservoir through hot springs to the surface (Chaudhuri et al., 2018; Dickson and Fanelli, 2003). It is noteworthy that the stable, inert gases such as He, Ar, and also the radioactive inert gases such as $^{222}$Rn, $^{220}$Rn are continuously migrating upward from the Earth's crust to the overlying atmosphere through diffusion and advection (Sathaye et al., 2016; Etiope et al., 2002; Zhao et al., 1998). This process, known as 'Earth degassing', is non-uniform (King, 1986; Damon and Kulp, 1958) in nature over space and time. The prominent signature of this degassing is usually seen along active faults, fractures, oceanic ridges, geothermal fields and even deep wells (King et al., 1996; Nagar et al., 1996; Wakita et al., 1980). Moreover, the ratio ($R=$) $^3$He/$^4$He in subterranean fluids and terrestrial matter is resolute by the proportional presence of two inherently different components of He- (a) primordial He (with $R \sim 10^{-4}$) which was seized within the earth during its accretion process and (b) radiogenic He (with $R \sim 10^{-8}$) that was produced within the earth due to radioactive disintegration of U and Th radio-nuclei. In atmospheric air, the value of $^3$He/$^4$He ($R$) is generally $1.4 \times 10^{-6}$ (Kennedy et al., 2000). However, any region where $^3$He/$^4$He ratio higher than ~10 times the atmospheric value is observed, is generally inferred as a tectonically active region along with incursion of a He plume from the lower mantle, even in the absence of other secondary evidence (Graham et al., 1993). $^3$He generated from the mantle is progressively diluted by mixing of $^4$He produced by means radioactive disintegration of the radio-elements in the earth's crust. The value of R at a given site measures the fastness of the fluid propagation to the surface from the underneath mantle (Denghong et al., 1999). Therefore, $^3$He/$^4$He ratios in terrestrial fluids are implicit indicators for the presence of magmatic components in the fluids and reflect the Physico-chemical process occurring at the deep earth. The said ratio is a remarkably sensitive variable relating to mantle-crust interactions (Craig et al., 2015). Moreover, $^3$He/$^4$He ratio of gases expelled through different terrestrial manifestation mirrors on the origin of He in the gases and is a key parameter to the classification and characterization of its sources.

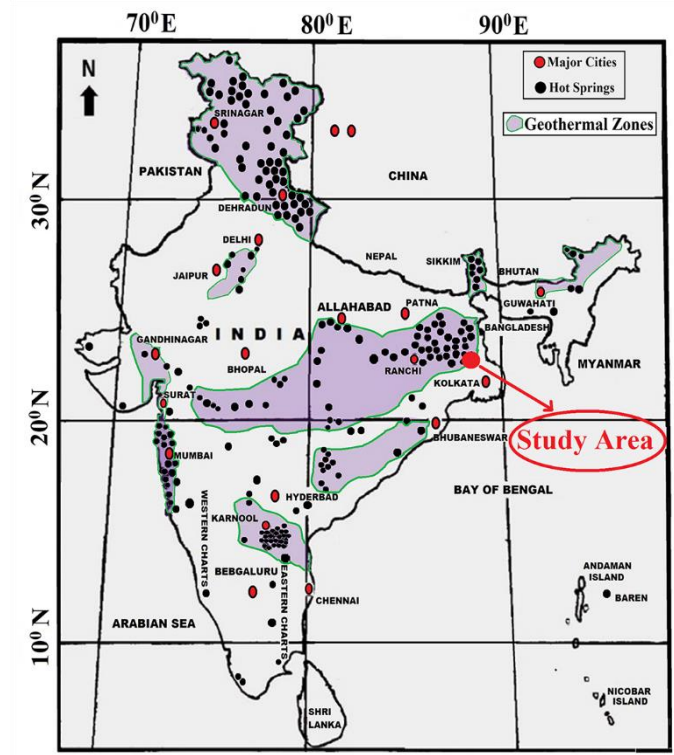

**Figure 1:** Location of the study area Bakreswar in the map of India (modified after Chaudhuri et al., 2018).

It is notable that geothermal energy sources are still overlooked in India for power generation even after the existence of a lot of potential sources, as seen in more than 300 hot springs throughout the country in the twelve geothermal zones of India (Chaudhuri et al., 2018). However, many of them could be well utilized for power generation using Earth's internal heat by means of developing geothermal power plants. For the sake of investigation, the hot spring site at Bakreswar in West Bengal, India, was selected as shown in Fig. 1. Several researchers have conducted several investigations on this geothermal area ever so often towards the exploitation of He and geothermal resources. Temperature and He emanation profile of some sites at Bakreswar geothermal provinces were recorded by Chaudhuri et al. (2015). Moreover, the audio-magnetotelluric (AMT) studies of the sub-surface beneath the Bakreswar geothermal area were conducted by Sinharay et al. (2010), as shown in Fig. 2. Now, knowing the amount of by-product, He gas which is ultimately reaching the surface through the fracture, fissure and hot springs vents etc., the associated heat energy (radiogenic) produced inside the reservoir can be estimated. The energy released per unit time from underneath bedrock at the study area was calculated by means of measuring the average amount of He emanated from Agni Kunda hot spring at Bakreswar. Here mainly the decay series of $^{238}$U, $^{235}$U and $^{232}$Th were considered and the amount of heat energy contributed due to each series was evaluated. Here the question may arise that each decay series (equation no. 1 to 3) takes a long period (in geological time scale) to complete its disintegration process and release of certain amount of heat in a discrete way, but heat generated due to each series are utilized to calculate the amount of heat production at the said reservoir at a certain instant of time. However, He emanation at the study area shows stable activities for a long-time-interval, as established by Chaudhuri et al. (2018). Therefore, He generation is also stabilized for a long period i.e., He generation due to every radioactive decay series and emanation of the said gas is in an equilibrium condition. Therefore, no He is being stored at the reservoir at the instant and therefore the He emanation could be considered to be equal to the generation of the same due to radioactive disintegration.

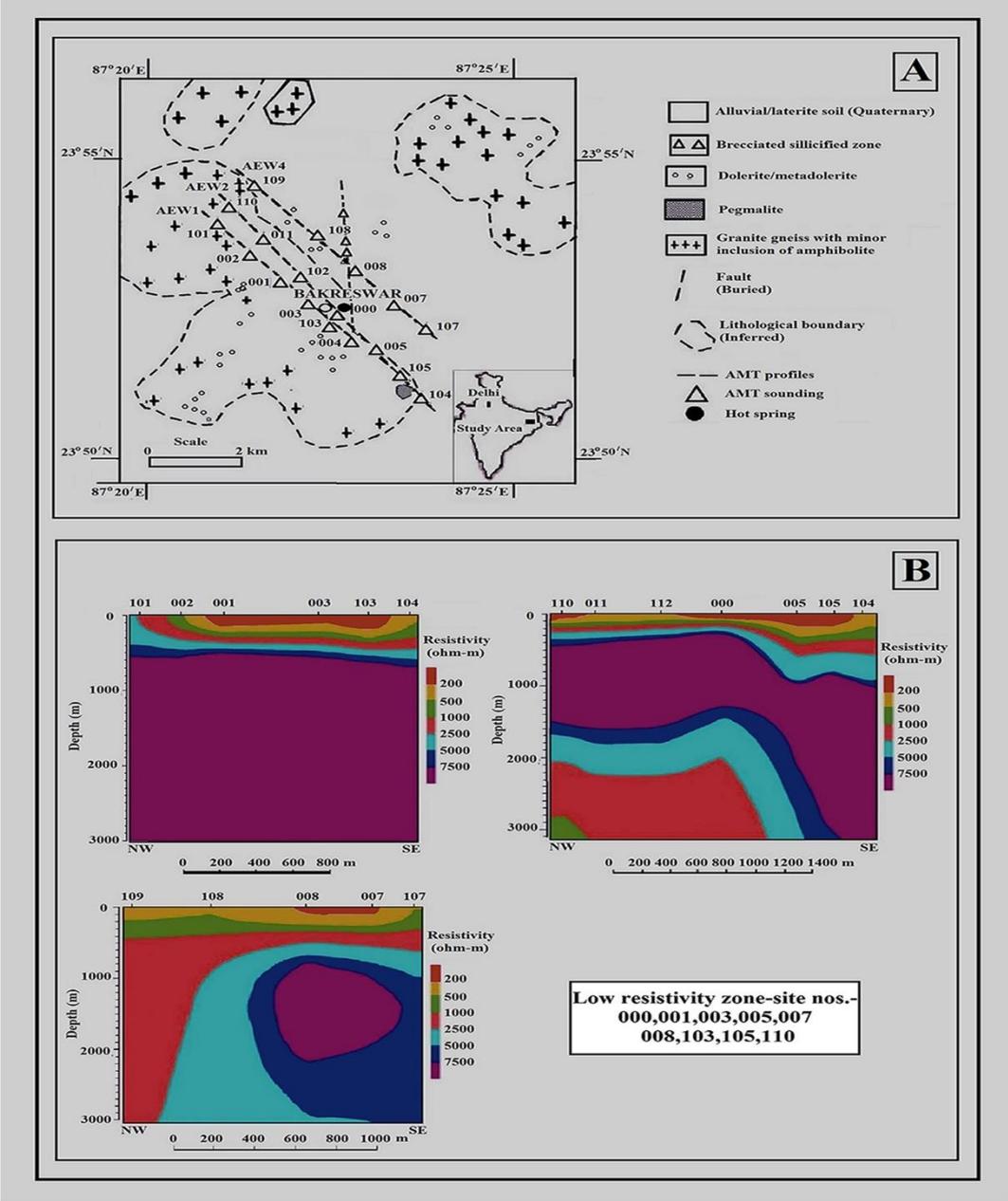

**Figure 2:** Resistivity profile at shallow depth at Bakreswar geothermal region (modified after Sinharay et al., 2010).

**Table I:** $^3$He/$^4$He ratio measurement at Bakreswar (geothermal area) and Kolkata (non-geothermal areas) (Chaudhuri et al., 2010).

| Sl. No. | Gas sample | Location | Geomorphology | Lat. °N; Long. °E | He Conc. (vol%) (GCA) | He$^3$/He$^4$ ratio (MSA) |
|---|---|---|---|---|---|---|
| 1 | Atmospheric Air (1 m height) | Kolkata, West Bengal | NGA (ambient temperature) | 22°33′32″ 88°20′24″ | 0.00052 | 1.37×10$^{-6}$ (normal) |
| 2 | Hot spring gas | Agni kunda hot spring, Bakreswar | GA (~69 °C) | 23°52′30″ 87°22′30″ | 1.48000 | 2.95×10$^{-6}$ (normal) |
| 3 | Hot spring gas | Agni kunda hot spring, Bakreswar | GA (~69 °C) | 23°52′30″ 87°22′30″ | 2.63000 | 5.89×10$^{-6}$ (Pre-seismic[a]) |
| 4 | Hot spring gas | Tantloi hot spring, Jharkhand | GA (~66 °C) | 24°01′00″ 87°17′00″ | 1.26000 | 0.91×10$^{-6}$ (normal) |

*Note: NGA = Non-geothermal Area; GA = Geothermal Area; Lat. = Latitude; Long. = Longitude; Conc. = Concentration; MSA = Mass spectrometer analysis; GCA = Gas Chromatography Analysis*

Moreover, $^3$He/$^4$He ratios were measured in hot spring gases at Bakreswar, West Bengal & Tantloi, Jharkhand (geothermal area) and in atmospheric air at Kolkata, West Bengal (non-geothermal area) using mass spectrometry analysis by Chaudhuri et al. (2010). In that investigation, $^3$He/$^4$He ratio (1.37×10$^{-6}$) of atmospheric air at Kolkata (220 km SE from Bakreswar) was taken as the running standard. $^3$He/$^4$He ratios of the collected (hot spring) gases from different geothermal provinces were compared against air standard. Two samples from each of the springs were collected within a fortnight during January 09, 2009 to February 10, 2009 (Chaudhuri et al., 2010) as shown in the Table I. The table reflects that $^3$He/$^4$He ratios for both the springs are higher than the standard values which signify high mixing rate of $^3$He rich magmatic gases into the reservoir of the geothermal system (Chaudhuri et al., 2010). But, implicit estimation of magmatic $^3$He incurred in the aquifer, and thus heat generated due to that is not possible to carry out separately due to the limitation of recorded data and technical difficulties. Moreover, the $^3$He/$^4$He ratio, as well as the He concentration, significantly increased during the pre-seismic event, implying Bakreswar geothermal provinces to fall under a tectonically active region (Chaudhuri et al., 2010).

---

[a] Pre-seismic refers the measurement time prior to any major earthquake recorded around the study area where as 'normal refers the same without any major earthquake afterwards.

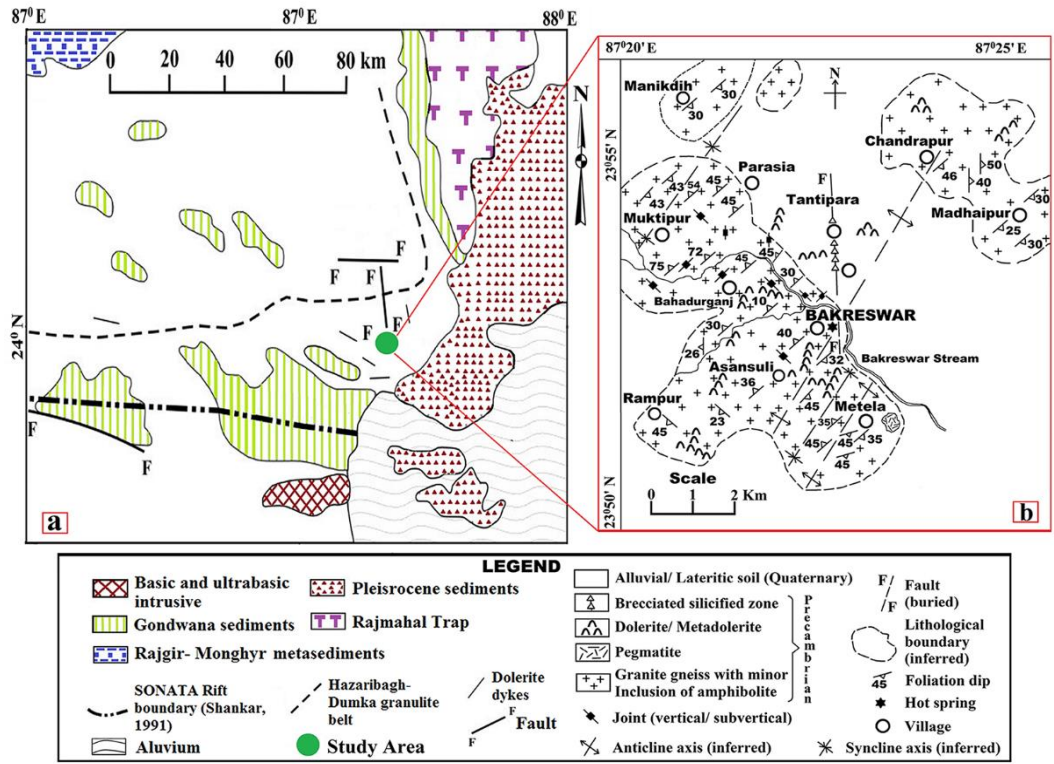

**Figure 3:** (a) Regional and (b) Local geological maps of the study area (modified after Nagar et al. 1996; Chaudhuri et al., 2018).

## 2. The study area: Bakreswar geothermal province

A group of seven[b] hot springs is scattered over Bakreswar geothermal area within a surface area of about 3350 sq. m (Chaudhuri et al., 2018 & 2015) and the area is located at the eastern end of the Son–Narmada–Tapi (SONATA) geothermal province which is a geologically complex, heterogeneous, and extensively faulted region (Fig. 3, window a) (Chaudhuri et al., 2018 & 2015). The area lies in the West Bengal Basin (WBB), which is the extension of the Chotanagpur Gneissic Complex (Ghose et al., 2002). Furthermore, it is associated with a 1.2-km-long shear zone characteristic by 50 m wide breccia/cherty quartzite trending in an almost

---

[b]He emanated from other six hot springs and through the soil (soil gas) of the geothermal area are not included in the estimation due to the lack of other necessary information.

N-S direction (Nagar et al., 1996) (Fig. 3, window b). The springs here are associated with the extinct (115 Ma) Rajmahal volcanic activity which is linked to Precambrian granitic rocks (Majumdar et al., 2000) (Fig. 3, window b). The highly sheared, brecciated, and mylonitized rock in this region facilitates the presence of highly porous and permeable subsurface of the Earth here (Nagar et al., 1996). Bakreswar geothermal area appears to be in a stressed state as it is associated with the eastern edges of two major fault systems—the SONATA fault and the ONGC fault (Majumdar et al., 2005; Shanker, 1991; Gupta et al., 1975). Over the geologic times, this area has been subjected to different cycles of plate movements along with intervening periods of isostatic rearrangements over the Precambrian and Cenozoic era (Shanker, 1991). This region exhibits a very high geothermal gradient (~ 90 °C/km) as well as a high heat flow rate (~ 230 mW/m2) (Chaudhuri et al., 2018; Shanker, 1988). In addition, electrical resistivity studies in this area confirm the presence of a high heat-conducting zone at a depth of about 2.8 km that goes down to 4 km (Shalivahan et al., 2004). This zone is supposed to act as a heat feeder to the fault system associated with the hot springs here. It is also noteworthy that Bakreswar hot springs are ascribed to the deep circulation of meteoric water along with the major fractures that have been recently created or reactivated in the basement crystallites in response to the tectonic disturbances (Majumdar et al., 2000). Moreover, the crustal thickness at Bakreswar is only 24 km whereas the average of same is 38 km in the rest of the country (Mukhopadhyay et al., 1986; Majumdar et al., 2000; Ghose et al., 2002). The average density of the crustal substance in this region is comparatively low. The thinner lithospheric overburden in this geothermal region allows easy transmission of inert volatiles such as He and $^{222}$Rn gases to impregnate through crustal constraints. As a result, the spring gases and the soil gases in the region are dominated by the presence of high $^{222}$Rn as well as He flux (Ghose et al., 2002). High $^{222}$Rn and He emanation are continuously conveyed and dispersed into the surrounding atmosphere through the formation of micro-bubbles at spring vents and molecular diffusion. In this connection, temperature and He emanation profile of some sites of Bakreswar geothermal provinces are also tabulated in Table II for a reference to attain the brief geophysical properties of the study area.

      Several researchers have investigated the hot springs group at Bakreswar ever so often. Chatterjee, S. D. (1972) showed that the high concentration of Ar at Agni Kunda hot spring at Bakreswar was contributed due to the dissolved Ar in sea-water, which was trapped within the bowels of the earth during the bygone age. Moreover, high radioactivity in the terrains of Bakreswar was likely to be of Triassic origin (Chatterjee, 1972). Vertical electrical sounding (VES) was also carried out in and around Bakreswar by Majumdar et al. (2000), and the

presence of a nearly N-S striking buried fault allowing passage for hot water to emerge in the form of springs was identified. The chemical composition of hot springs examined by Mukhopadhyay and Sarolkar (2012) indicated that the spring water was of meteoric origin. The same was also identified by Majumdar et al. (2005) by means of observing the seasonal variations in the isotopes of $O_2$ and H in geothermal waters from Bakreswar and recharge of the springs was expected to take place somewhere in the vicinity of Gondwana basins. Majumdar et al. (2009) estimated the possible temperature of the reservoir at a depth of about 1 km to be 100±5 °C. The reservoir temperature at the study area was estimated to be in the range of 130 °C to 175 °C (by Na /K ratio) and 110 °C to 124 °C (by $TSiO_2$) (Mukhopadhyay and Sarolkar 2012). Furthermore, the range of reservoir temperature was also calculated as 212 °C to 124 °C, 118 °C to 120 °C and 126 °C to 130 °C by means of using silica geothermometry by respectively.

**Table II:** Temperature and He emanation profile of some sites at Bakreswar geothermal provinces.

| Sl. No. | Test Site (distance from Agni Kunda) | Sample type | Temperature (°C) | He Conc. (vol %) | References |
|---|---|---|---|---|---|
| 1 | Bakreswar Agni Kunda (0 m) | HSG | 69.0 | 1.72 | Chaudhuri et al., 2018 |
| 2 | Bakreswar Khar Kunda (16 m) | HSG | 68.0 | 1.36 | Chaudhuri et al., 2010 |
| 3 | Bakreswar Bhairab Kunda (7 m) | HSG | 62.0 | 1.12 | Chaudhuri et al., 2015 |
| 4 | Bakreswar Brahma Kunda (20 m) | HSG | 46.0 | 1.26 | Chaudhuri et al., 2015 |
| 5 | Bakreswar Surya Kunda (18 m) | HSG | 63.0 | 0.31 | Chaudhuri et al., 2015 |
| 6 | Bakreswar Reserve Tank (5 m) | HSG | 52.0 | 0.91 | Chaudhuri et al., 2015 |
| 7 | Bakreswar PWD Bungalow site (987 m) | SG (1 m depth) | 32.0 (Ambient) | 0.35 | Chaudhuri et al., 2019 |
| 8 | Bakreswar PWD Bungalow site (988 m) | SG (3 m depth) | 31.0 (Ambient) | 0.02 | Chaudhuri et al., 2019 |
| 9 | Bakreswar PWD Bungalow site (990) | AA (1 m height) | 33.0 (Ambient) | 0.05 | Ghose et al., 2002 |
| 10 | Bhabanipur (10 km) | AA (1 m height) | 28.0 (Ambient) | 0.07 | Ghose et al. 2002 |
| 11 | Mallarpur (43 km) | BG (100 m depth) | 58.0 | 1.20 | RWA |

*Note: HSG = Hot spring gas; SG = Soil gas; AA = Ambient air; BG = Borehole gas; Conc. = Concentration; RWA = Recent work by the authors*

### 3. Methodology

#### 3.1. Experimental techniques

A field laboratory was established at the hot spring site of Bakreswar for continuous monitoring of gases coming out from the hot spring Agni Kunda. A giant inverted SS funnel was placed underwater in the hot spring (Agni Kunda) at a position where gas out flux was significantly high, to trap hot spring gases which were comprised of He, Ar, $O_2$, $N_2$, $CH_4$, $CO_2$, $^{222}Rn$, etc. A programmable and portable micro gas chromatograph CP 490 (make Agilent) provided with a micro thermal conductivity detector, was utilized to detect the relative concentration of different gases present in the spring gas. Here ultra-pure (>99.998 vol%) hydrogen gas was used as the carrier gas for running the equipment. The entire measurement was carried out for a continuous five-year (August 01, 2005 to July 31, 2010) in around the clock (24×7) measurement fashion. The back-up power supply was maintained to keep a continuous and stable power supply in case of a power failure. The schematic diagram of the experimental set up is shown in Fig. 4. Details of the aforesaid experimental procedure are already described elsewhere (Chaudhuri et al., 2018). The average value of the He concentration (vol%) of 5 years of continuous measurement was adopted in our study. Moreover, the flow rate of the emanated gases from the spring vent was measured by means of collecting the spring gases in a gas container (5 litre) from the main channel of the incoming gas line. The gas collection procedure was kept running up to a certain time until its pressure makes equilibrium with the pressure (1.58 atm) at the spring vent underwater. This type of measurement was done once every month for a continuous five years, and the average value of those was considered as the final value of the gas flow rate under consideration. The ambient temperature of the study area was monitored for the same interval at the time of measurement of gas flow rates.

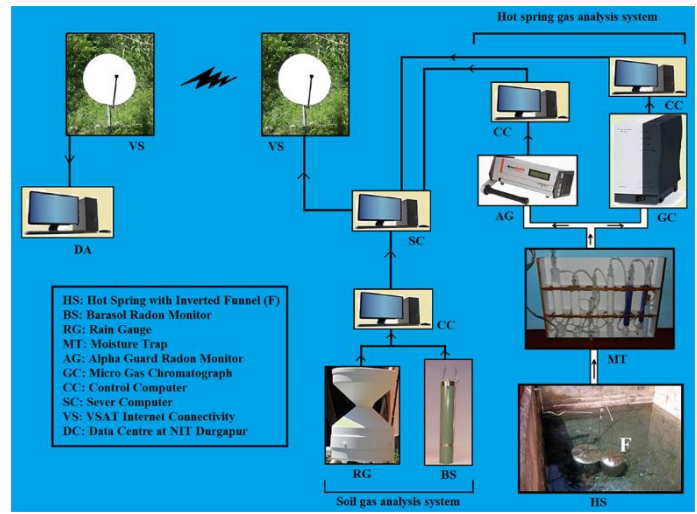

**Figure 4:** Schematic diagram of the experimental set up installed at the field site Bakreswar (modified after Chaudhuri et al., 2018).

### 3.2. Analytical techniques

To move towards the desired direction for calculation, the following steps were adopted.

The volume of He gas ($V_{He}$) emanating from the spring per second was estimated as

$$V_{He} = F \times C_{He} \qquad (4)$$

Where $F$ = average flow rate of He gas emanation (recorded); $C_{He}$ = relative concentration of He in the gas mixture, which was expelled through the spring vent (recorded). The no. of moles of He gas emanated per unit second from the hot spring was calculated using the real gas equation as stated below:

$$(V - nb)\left(P + \frac{n^2 a}{V^2}\right) = nRT \tag{5A}$$

$$\text{i.e., } \frac{ab}{V^2}n^3 - \frac{a}{V}n^2 + (bP + RT)n - PV = 0 \tag{5B}$$

Where $V$ (i.e. $V_{He}$) = volume of the He gas at temperature T and pressure P; R= universal gas constant= 0.0821 litre atm/ mole K; $n$= number of mole (to be calculated); 'a' and 'b' are real gas constants and for He, a= 0.03457 atm litre$^2$/ mole$^2$; b= 0.02370 litre/mole (Fishbane et al., 2005). Solving the equation no. 5B and considering the real root for 'n', the corresponding total number of He atom ($N_{He}$) was calculated by

$$N_{He} = n \times N_A \tag{6}$$

Here, $N_A$= Avogadro's number = $6.022140857 \times 10^{23}$ (Fox and Hill, 2017). The relative contribution of the individual isotope in the generation of He atoms was calculated according to their relative abundance in the natural resources because the total number of He atom is produced via the radioactive decay series of $^{238}$U, $^{235}$U, and $^{238}$Th. Here, the same was not applicable to the $^{40}$K series as it disintegrates through only β emission. Therefore, for production of He atoms,

$$\text{The relative contribution of Uranium } (C_U) = \frac{\text{conc. of Uranium (U)}}{\text{total conc. of Uranium (U) \& Thorium(Th)}} \tag{7}$$

$$\text{The relative contribution of Thorium } (C_{Th}) = \frac{\text{conc. of thorium}}{\text{total conc. of Uranium (U) and Thorium(U)}} \tag{8}$$

The basement of the study area is predominantly composed of granite gneiss belonging to the Precambrian Chotanagpur Gneissic Complex (Nagar et al., 1996; Majumdar et al., 2000). Here the relative contribution of U and Th were evaluated according to their (average) content in granite type rock material i.e. $^{238}$U [or $^{235}$U] content as 4.80 ppm and $^{232}$Th content as 21.50 ppm were considered (Pereira, 1980; Brown, 2010). Moreover, natural Uranium is an admixture of $^{238}$U (99.28%) and $^{235}$U (0.71%) (Pereira, 1980). Therefore, for production of He atoms by radioactive decay,

$$\text{The relative contribution of }^{238}\text{U } (C_{U-238}) = \frac{99.28}{100} \times C_U \tag{9A}$$

$$\text{The relative contribution of }^{235}\text{U } (C_{U-235}) = \frac{0.71}{100} \times C_U \tag{9B}$$

The no. of the He atoms generated (in a unit second) due to the decay of radio nuclei $^{238}$Th, $^{238}$U and $^{235}$U are respectively $C_{Th-232} \times N_{He}$, $C_{U-238} \times N_{He}$ and $C_{U-235} \times N_{He}$. According to the equation no. 1 to 3, it is reflected that 6 He atom and 42.6 MeV/atoms heat energy, 8 He atoms and 51.7 MeV/atom heat energy and 7 He atom and 46.4 MeV/atoms heat energy from the decay of $^{238}$Th, $^{238}$U and $^{235}$U are releasing respectively. Therefore, the energy release (per unit second) from $^{238}$Th, $^{238}$U, and $^{235}$U decay can be evaluated respectively by

$$E_{Th-232} = \frac{42.6}{6} \times C_{Th-232} \times N_{He} \quad (10)$$

$$E_{U-238} = \frac{51.6}{8} \times C_{U-238} \times N_{He} \quad (11)$$

$$E_{U-235} = \frac{46.4}{7} \times C_{U-235} \times N_{He} \quad (12)$$

And the total energy generated due to the decay of all these three radioelements were

$$E_{R(Total)} = E_{Th-232} + E_{U-238} + E_{U-235} \quad (13A)$$

An important issue to discuss is that the loss of generated heat energy may be considered to be negligible here as a capping of the impermeable and insulating bedrock over the geothermal system prevents the heat transfer by means of conduction and convection (Sinharay et al., 2010; Majumdar et al., 2000; Majumdar, 2010). Therefore, the heat energy would be stored inside the geothermal system which may be subjected to break its dynamical stability after the accumulation of enough energy within it. However, that doesn't happen as excess heat is drained to the surface along with the transfer of geothermal fluid through the spring vent (Majumdar et al., 2000; Majumdar, 2010). Moreover, here only radiogenic heat is accounted for and the contribution of energy belonging to primordial heat sources is not included. However, Gando et al. (2011) (Gando et al., 2011) documented that heat from radioactive decay was contributed about half of Earth's total heat flux and the rest was accounted from the primordial heat source of the Earth. Considering the similar concept, we can also assume that the primordial heat source also would contribute as much as heat energy generated by radioactive decay of radio-nuclei at the reservoir of the study area.

Therefore, Heat generated by the primordial source,

$$E_{P(Total)} = E_{R(Total)} \quad (13B)$$

Therefore, total energy contributed from the radiogenic and primordial source is

$$E_{Total} = E_{P(Total)} + E_{R(Total)} \quad (14)$$

Moreover, If the geothermal gradient $\left(\frac{d\theta}{dx}\right)$ is considered to be constant at least up to the depth of the reservoir (x) then the depth (x) of the reservoir could be calculated from the below stated linear relationship

$$\theta_r = \left(\frac{d\theta}{dx}\right)x + \theta_a \quad (15)$$

Where, $\theta_r$ = reservoir temperature of the geothermal system; $\theta_a$ = average ambient temperature at the study area.

## 4. Result and discussion

All the calculated values such as n, $N_{He}$, $C_{Th}$, $C_{U-238}$, $C_{U-235}$, $E_{Th-232}$, $E_{U-238}$, $E_{U-235}$ and $E_{Total}$ are listed in Table III. The heat energy generated per unit second by the radioactive decay of $^{238}$Th, $^{238}$U and $^{235}$U inside the reservoir were calculated as 31.58 MW, 6.3585 MW and 0.0467 MW respectively and together contributed as approximately 38 MW (radiogenic source). Combining radiogenic and primordial source, heat generation was supposed to be 76 MW.

**Table III:** Experimental and calculated parameters.

| Sl. No. | Parameters | Parameters' value | References (equation no., if any) |
|---|---|---|---|
| 1 | He concentration, $C_{He}$ | 1.72 vol% | RA |
| 2 | Flow rate, F | 3.5 L/min | RA |
| 3 | He emanation per minute, $V_{He}$ | 0.0602 L/min | EA (4) |
| 4 | Temperature inside the spring gas trapping funnel, T | 342 K (69 °C) | Chaudhuri et al., 2018 |
| 5 | Pressure inside the spring gas trapping funnel, P | 1.58 atm | RA |
| 6 | Number of moles emanating per second, n | $56.3877 \times 10^{-6}$ | EA (5B) |
| 7 | Total number of He atoms emanating per second, $N_{He}$ | $33.9623 \times 10^{18}$ | EA (6) |
| 8 | The relative concentration of $^{232}$Th, $C_{Th}$ | $81.7490 \times 10^{-2}$ | EA (8) |
| 9 | The relative contribution of $^{238}$U, $C_{U-238}$ | $18.1195 \times 10^{-2}$ | EA (9A) |
| 10 | The relative contribution of $^{235}$U, $C_{U-235}$ | $0.1296 \times 10^{-2}$ | EA (9B) |
| 11 | Energy contributed due to decay of $^{232}$Th, $E_{Th-232}$ | 31.58 MW | EA (10) |
| 12 | Energy contributed due to decay of $^{238}$U, $E_{U-238}$ | 6.3585 MW | EA (11) |
| 13 | Energy contributed due to decay of $^{235}$U, $E_{U-235}$ | 0.0467 MW | EA (12) |
| 14 | Energy accounted for radiogenic source, $E_{R(Total)}$ | 37.9834 MW | EA (13A) |
| 15 | Energy accounted for primordial source, $E_{P(Total)}$ | 37.9834 MW | EA (13B) |
| 16 | Total Energy accounted from radiogenic & primordial source, $E_{Total}$ | 75.9668 MW | EA (14) |
| 17 | Geothermal gradient, $\frac{d\theta}{dx}$ | 90 °C/km | Shanker, 1988 |
| 18 | Reservoir temperature, $\theta_r$ | 130 °C | Gupta et al., 2016 |
| 19 | Average ambient temperature, $\theta_a$ | 26 °C | RA |
| 20 | Depth of the geothermal reservoir, x | 1155 m | EA (15) |

*Note: EA = Estimated by the authors; RA = Recorded by the authors*

Recently, the reservoir temperature of Bakreswar geothermal system was estimated to be 126-130 °C by Gupta et al. (2016) by means of silica geothermometry. The average ambient temperature of the study area was recorded to be 26 °C. Our estimation using the mathematical relation (equation no. 15) shows that the geothermal reservoir at the area is expected to be located at about 1,111 to 1,155 m beneath the surface. Moreover, the audio magnetotelluric (AMT) studies revealed the existence of a deep heat reservoir in the N–W part of Bakreswar. Observation sites 000, 001, 003, 005, 007, 008, 103, 105, 110 as marked in Fig. 2 show low resistivity profile and these are more favorable sites for deep drilling.

## 5. Conclusion

Using a simple technique by means of He exploration study at the field site, the probable energy generated inside the reservoir was estimated here. Considering the combined source of heat generation inside the reservoir system, the energy was expected to be harness from the source of power of 38 to 76 MW using the appropriate technology. Moreover, the values would be likely increased whenever, the He emanations through the others hot springs where He emanation is comparably less than that of Agni kunda) and through the vast surface area (soil gas) at Bakreswar would be included in this estimation. However, this was a little bit difficult as well as complicated due to the technical coerces and geographical constraints. Furthermore, the value of the said power may increase when deep drilling would be made at a location near to the hot spring area as indicated in Fig. 2. It is notable that Kalina cycle based binary power plant using Ammonia–water mixture as working fluid (thermal efficiency: 13–53%) (DiPippo, 2004), is already proposed to be installed at the spring site (Das et al., 2016). Accordingly, if such type of power plant is supposed to be installed, for say, the plant would be capable to deliver the power of 4.94 MW (minimum thermal efficiency 13%) to 20.13 MW (maximum thermal efficiency 53%) only considering the radiogenic heat source. These values are changed to 9.88 MW and 40.26 MW respectively when the primordial heat source is comprised of the radiogenic heat source.

**Acknowledgment**

The authors owe a debt of gratitude to the National Institute of Technology Durgapur (NIT Durgapur) and the Ministry of Human Resource Development (MHRD), Govt. of India for providing the financial as well as institutional support in all respect for carrying out such type of the research activities at the field site, Bakreswar.

Mburu, M., "Geothermal Energy Utilisation. Presented at Short Course IX on Exploration for Geothermal Resources," UNU–GTP, GDC and KenGen at Lake Bogoria and Lake Naivasha, Kenya (2009). Available online at:

http://www.unugtp.is/en/moya/page/sc-19, accessed on February 06, 2020

Mukhopadhyay, M., Verma, R. K., Ashraf, M. H. "Gravity field and structures of the Rajmahal Hills: example of the paleo-Mesozoic continental margin in eastern India," Tectonophysics **131(3)**, 353–367 (1986).

Mukhopadhyay DK, Sarolkar PB. "Geochemical appraisal of Bakreshwar-tantloi hot springs, West Bengal and Jharkhand, India," Proc of 37th Workshop on Geotherm Reserv Eng, Stanford University, California (2012). Available online at:

https://pangea.stanford.edu/ERE/pdf/IGAstandard/SGW/2012/Mukhopadhyay.pdf, accessed February 06, 2020.

Nagar, R. K., Vishwanathan, G., Sagar, S., Sankaranarayanan, A., "Geological, geophysical and geochemical investigations in Bakreswar-Tantloi thermal field, Birbhum and Santhal Parganas Districts, West Bengal and Bihar, India," Geotherm. Energy in India, GSI Sp Pub 45, (In Pitale, U. L. and Padhi, R. N., eds.), GSI, India (1996), pp. 87-98.

Nicholson, K. "Geothermal Fluid Chemistry and Exploration Techniques," SpringerVerlag, Heidelberg, Germany (1993).

Pereira, E. B., "Some problems concerning the migration and distribution of He–4 and Radon–222 in the upper sediments of the crust–a theoretical model; and the development of a quadrupole ion filter for measuring He at the soil air interface," Ph. D Thesis, Rice University (1980), pp. 8. Available online at:

https://scholarship.rice.edu/handle/1911/15577, accessed on February 06, 2020

Sathaye, K. J., Smye, A. J., Jordan, J. S., Hesse, M. A., "Noble gases preserve history of retentive continental crust in the bravo dome natural $CO_2$ field, New Mexico," Earth Planet Sci. Lett. **443**, 32–40 (2016).

Shalivahan, S., Sinharay, R. K. & Bhattacharya, B. B. "Electrical conductivity structure over geothermal province of Bakreswar, Eastern India," Proc of 17th IAGA WG 1.2 Workshop on Electromagnet Induc in the Earth 2004, 18–23 (2004).

Shanker, R., "Heat-flow of India and discussion on its geological and economic significance", Indian Miner **42**, 89–110 (1988).

Shanker, R. "Thermal and crustal structure of SONATA—azoneof mid continental rifting in Indian shield," J Geol Soc India **37(3)**, 211–220 (1991).

Sinharay, R. K., Shalivahan S., Bhattacharya, B., "Audiomagnetotelluric studies to trace the hydrological system of thermal fluid flow of Bakreswar Hot Spring, Eastern India: A case history (Personal Communication)" (2016).

Wakita, H., Nakamura, Y., Kita, I., Fujii, N., Notsu, K., "Hydrogen release: new Indicator of fault activity," Sci. **210**, 188–190 (1980).

Zhao, X., Fritzel, T. L. B., Quinodoz, H. A. M., Bethke, C. M., Torgersen, T., "Controls on the distribution and isotopic composition of helium in deep ground-water flows," Geology **26(4)**, 291–294 (1998).17